# Tipping the Balance: a criticality perspective

Indrani Bose

Department of Physics

Bose Institute

93/1, A. P. C. Road

Kolkata-700009, India

Correspondence: ibose1951@gmail.com

Abstract: Cell populations are often characterised by phenotypic heterogeneity in the form of two distinct subpopulations. We consider a model of tumour cells consisting of two subpopulations: non-cancer promoting (NCP) and cancer-promoting (CP). Under steady state conditions, the model has similarities with a well-known model of population genetics which exhibits a purely noise-induced transition from unimodality to bimodality at a critical value of the noise intensity $\sigma^2$. The noise is associated with a parameter λ representing the system-environment coupling. In the case of the tumour model, λ has a natural interpretation in terms of the tissue microenvironment which has considerable influence on the phenotypic composition of the tumour. Oncogenic transformations give rise to considerable fluctuations in the parameter. We compute the $\lambda - \sigma^2$ phase diagram in a stochastic setting, drawing analogies between bifurcations and phase transitions. In the region of bimodality, a transition from a state of balance to a state of dominance, in terms of the competing subpopulations, occurs at λ = 0. Away from this point, the NCP (CP) subpopulation becomes dominant as λ changes towards positive (negative) values. The variance of the steady state probability density function as well as two entropic measures provide characteristic signatures at the transtion point.

1. Introduction

The generation of heterogeneity in a population is often a consequence of the underlying stochastic nonlinear dynamics. Cell biology offers a large number of examples of population heterogeneity characterized by the coexistence of multiple subpopulations with distinct phenotypic traits [1, 2, 3]. The frequently observed case is that of bimodality with two distinct subpopulations defining the heterogeneity. A well-understood physical basis of such bimodality is as follows. A single positive feedback loop or multiple loops governing the underlying deterministic dynamics create the potential for bistabiity in a specific parameter regime [4]. In the state space defined by the concentrations of the key dynamical variables, the two stable steady states are separated by an unstable steady state. The corresponding landscape picture is that of two valleys separated by a hill. The minima of the valleys define the stable steady states, the attractors of the dynamics, and the top of the hill the unstable steady state. The stochastic component of the dynamics generates fluctuations (noise) in the magnitudes of the dynamical variables opening up the possibility of fluctuation-driven transitions from one attractor in the state space to the other. In the landscape analogy, a ball, left to itself, resides at the bottom of one of the valleys. Noisy dynamics, in the form of, say, random kicks imparted to the ball, give rise to a finite probability that the ball crosses the top of the hill and reaches the other valley. The stochastic dynamics smear out the steady states in the form of a steady state probability distribution with two distinct peaks, the case of bimodality. In this case, bistable deterministic dynamics are essential for the observation of bimodality in the stochastic case.

In an alternative scenario, the bimodality is purely noise-induced with the deterministic dynamics yielding single stable steady states in the full parameter regime, even in the presence of positive feedback loops. In cell biology, there is now experimental evidence of noise-induced bimodality in the key cellular phenomenon of gene expression [5]. One of the earliest examples of purely noise-induced transitions, resulting in bimodality, relates to a model of population genetics, referred to as the genetic model [6,7]. In this model, the deterministic dynamics are governed by the rate equation

$$\frac{dx}{dt} = 0.5 - x + \beta x(1-x) \qquad (1)$$

in which $x$ represents a state variable, with $x \in [0,1]$, and $\beta$ is a parameter describing the coupling of the system to the environment with $\beta \in [-\infty, +\infty]$. The model has a realistic interpretation in a population genetics context as well as in the case of a scheme of chemical reactions [7]. From (1), the steady state is given by

$$x_s = (2\beta)^{-1}[\beta - 1 + \sqrt{1+\beta^2}] \qquad (2)$$

A globally stable unique steady state exists for each value of the external parameter $\beta$. In the presence of a fluctuating environment, the coupling parameter $\beta$ acquires the form, $\beta \rightarrow \beta + \sigma \xi$, where $\xi$ represents a white noise and $\sigma^2$ is the noise intensity. The dynamics now acquire a stochastic character requiring analysis in the framework of stochastic formalisms. The noise intensity serves as another parameter in the system, besides $\beta$, and one can show that a noise-induced transition to a bimodal steady state probability distribution occurs for $\sigma^2 > \sigma_c^2$, a critical noise intensity. One can identify a critical point at $\beta_c = 0, \sigma_c^2 = 2, x_s = 0.5$. For $\sigma^2 < \sigma_c^2$, the probability distribution has a single peak centred on $x_s = 0.5$. The single peak splits into a pair of peaks above the critical point. As in the case of equilibrium critical behavior, the noise-induced nonequiibrium transition at the critical point is characterized by a set of critical exponents. The critical behavior has been shown to belong to the mean-field Ising universality class [7]. For $\beta \neq 0$, the critical value of the noise intensity, above which bimodality is observed, becomes a function of $\beta$.

A recent study has shown that biochemical models with positive feedback also exhibit mean-field Ising-like critical behavior [8]. The transition to a bimodal distribution is, however, not purely noise-induced if the underlying deterministic dynamics give rise to bistability in a specific parameter regime. In general dynamical models, the dynamics undergo regime changes at the bifurcation points which are analogous to phase transition points in equilibrium models [9,10,11]. The bifurcations bringing about discontinuous (continuous) changes at the bifurcation points are akin to first-order (continuous/critical-point) phase transitions. In dynamical systems, the transition at a saddle-node (SN) bifurcation is discontinuous whereas that occurring at a supercritical pitchfork (SP) bifurcation is continuous. The latter bifurcation point has the character of a critical point. The mapping between the mean-field Ising and the nonequilibrium models rests on the equivalent forms of the equation of state and the steady state equation in the vicinity of the respective critical points. The effective thermodynamic quantities (temperature, magnetic field and magnetization) of the equilibrium model are expressed in terms of the effective biochemical parameters of the nonequilibrium model [8,10]. The utility of such mappings has been demonstrated in single cell experiments on T cell signaling [8].

A prominent example of population heterogeneity is that of intra-tumour heterogeneity (ITH) in terms of both genotypic and phenotypic variability [12,13]. Genotypic changes in the form of mutations are responsible for the formation of tumours, an abnormal conglomeration of cells and tissues. A tumour further harbours subpopulations with heterogeneity at both the genotypic and phenotypic levels. The first type of heterogeneity has been extensively studied whereas the interest in analyzing the second type of heterogeneity is of more recent origin. The study of phenotypic heterogeneity opens up the possibiity for developing targeted therapies against the subpopulations with a dominant role in the growth and spread of cancer.

Phenotypic diversity can have both genotypic and non-genetic origins. There is now considerable experimental evidence that the cells in a tumour are broadly of two types: cancer stem cells (CSCs) and non-cancer stem cells (NCSCs) [12]. The CSCs arise when normal stem cells undergo mutations. The two subpopulations have considerable phenotypic differences, e.g., the proliferative potential of the CSC is much higher than that of the NCSC. The protein HER2 ( human epithelial growth factor receptor 2) promotes the quick growth of cancer cells in a type of breast cancer [14, 15]. The tumour population has two phenotypically distinct subpopulations designated as HER2+ and HER2-. The former (latter) subpopuation has a higher (lower) HER2 protein level in a cell than the level prevailing in a normal cell. The HER2+ subpopulation has a higher growth rate and tends to spread (metastatize) at a faster rate. The dominance of a cancer-promoting subpopulation in a tumour tilts the balance towards the eventual fate, cancer.

The onset of cancer involves a transition from the healthy to the pre-disease state followed by a sudden deterioration to the disease state with its characteristic clinical symptoms [16,17]. In theoretical models, the sudden regime shift occurs via a saddle-node bifurcation so that the transition has the character of a first-order phase transition. The variation of an appropriate parameter generates a line of first order phase transitions terminating at a critical point in the phase diagram. A number of quantitative measures have been proposed to provide the early warning signals of an imminent transition to the pre-disease state [18,19,20]. Before this state is reached, one can reverse the progressive deterioration by adopting suitable measures like drug treatment. The sudden regime shift to the disease state brings in irreversibility due to the existence of hysteresis in the response curve. In this paper, we explore the critical-like behavior in a model of cancer in which the ITH is defined in terms of two distinct subpopulations. In Section 2, we introduce the model and show that it can be mapped onto the population genetics model with the dynamics as described in (1). The stochastic dynamics of the model are described using the formalism of the Fokker-Planck equation (FPE). The existence of a purely noise induced transition to bimodality at a critical point is demonstrated and the critical exponents are identified as those of the mean-field Ising type. In Section 3, the phase diagram in the $\lambda - \sigma^2$ plane is computed with the phase boundaries separating regions of unimodality from that of bimodality. The computation is based on stochastic considerations arising from the formalism of the FPE. The phase diagram is interpreted in terms of the sudden regime shifts to the disease state. The corresponding hysteresis curve is also exhibited. In Section 4, the transition from the regime in which both the subpopulations have equal probability for dominance to a regime in which one of the subpopulations is dominant, is investigated. The signatures of the transition, which tips the balance towards a specific subpopulation, are obtained via some statistical measures. Section 5 contains some concluding remarks.

1. Model of Tumour Heterogeneity
1.1 Similarity with Population Genetics Model

We consider a tumour population the heterogeneity of which is defined in terms of two distinct subpopulations, say, HER2+ and HER2- subpopulations or cancer-promoting (CP) and non-cancer-promoting (NCP) subpopulations. Let $n_1(t)$ and $n_2(t)$ be the sizes (size is given by number of cells) of the NCP and CP subpopulations respectively at time $t$ with the total population size $n(t) = n_1(t) + n_2(t)$. Each cell in a subpopulation can undergo a symmetric cell division to yield a pair of daughter cells belonging to the same subpopulation resulting in the growth of the subpopulation. Let $k_1$ and $k_2$ be the growth rate constants associated with the NCP and CP subpopulations respectively. These rate constants define effective growth rates when the cell death due to apoptosis is included in the rates. A cell in a subpopulation can also divide asymmetrically with one daughter cell belonging to the parent subpopulation and its companion to the other subpopulation. Through this process of interconversion, the other subpopulation increases in size [14] whereas the size of the parent subpopulation remains the same. The rate constants of these processes are $r_1$ (asymmetric division of a cell in the NCP subpopulation) and $r_2$ (asymmetric division of a cell in the CP subpopulation). The dynamical equations capturing the above described processes are [15]:

$$\frac{dn_1}{dt} = k_1 n_1(t) + r_2 n_2(t) \quad (3)$$

$$\frac{dn_2}{dt} = k_2 n_2(t) + r_1 n_1(t) \quad (4)$$

The symmetric division event giving rise to two daughter cells belonging to the other subpopulation are not considered explicitly as such events are of rare occurrence [15]. The dynamical equations in (3) and (4) are similar in form to those appearing in a number of studies dealing with the problem of phenotypic heterogeneity [15, 21, 22, 23]. We now define $f_1(t)$ and $f_2(t)$ to be the fractions of the total population belonging to the NCP and CP subpopulations respectively at time $t$ with $f_1(t) = \frac{n_1(t)}{n(t)}$ and $f_2(t) = 1 - f_1(t)$. Expressing $f_2(t)$ in terms of $f_1(t)$, (3) and (4) can be combined into a single equation for $f_1$, namely

$$\frac{df_1(t)}{dt} = k_{diff} f_1(t)(1 - f_1(t)) + r_e - 2r_e f_1(t) \quad (5)$$

where $k_{diff} = k_1 - k_2$ and the rate constants $r_1$ and $r_2$ are taken to be equal [15], $r_1 = r_2 = r_e$. The theoretical results, for the evolution of the subpopulation fractions $f_1$ and $f_2$ as a function of time and with different initial conditions, fit the experimental data quite well [15] conferring validity on the tumour model. Through a simple rescaling of the time, $t_D = 2r_e t$, and introducing the dimensionless parameter, $\lambda = \frac{k_{diff}}{2r_e}$, (5) is rewritten in the form

$$\frac{df_1}{dt_D} = \lambda f_1(t) \left(1 - f_1(t)\right) + \frac{1}{2} - f_1(t) \quad (6)$$

We point out that the dynamical equation (6) has the same form as that of the population genetics model [6,7]. One can draw parallels between the two models by noting that the population genetics model considers a constant population of haploid individuals in which two alleles, with frequencies $x$ and $1 - x$ in the population, compete for a specific gene locus. The frequencies undergo changes due to two mechanisms: mutation between the two alleles and natural selection. The parameter β in (1) is a selection coefficient with dependence on the state of the environment. A positive (negative) value of β favours the allele with frequency $x$ ($(1-x)$). The term $\frac{1}{2} - x$ in (1) represents the change in frequency brought about by random mutations. In the cancer model, described by (3) and (4), the two mechanisms for the change in frequency are the two-way transitions between the phenotypes, analogous to mutations, and the unequal growth rates of the subpopulations, favouring one subpopulation over the other, as in natural selection. The crucial difference between the two models is that in the genetic model, the total population size is consevred whereas in the case of the tumour model the total size $n(t)$ is a function of time. The dynamical variables, $f_1$ and $f_2$, representing the fractions of the total population that belong to the NCP and CP subpopulations respectively, can, however, attain finite steady state values due to a cancellation of the time-dependent factors in the numerator and the denominator of $f_1 = \frac{n_1(t)}{n(t)}$ and $f_2 = \frac{n_2(t)}{n(t)}$, in the limit of large times (attainment of the steady state). This is possible due to the process of interconversion between the phenotypes via asymmetric cell division. In Appendix A, we show this explicitly for the tumour model (with dynamics as in (3) and (4)) under steady state conditions. The steady state value of $f_1$ is given by the expression in (2) with λ replacing β, confirming the validity of the reduced tumour model in (6). From the general to the reduced tumour model, the number of parameters decreases from four to one.

### 1.2 Critical-point Transition to Bimodality

Taking cues from the population genetics model [6, 7], the results for a critical point transition to bimodality in the cancer model are as follows. The stochastic dynamics are governed by the differential rate equation:

$$\frac{dq}{dt} = \frac{1}{2} - q + \lambda q(q-1) + \sigma g(q)\xi(t) \qquad (7)$$

where $q$ and $1-q$ represent the fractions of the total population belonging to the NCP and CP subpopulations respectively and $g(q) = q(1-q)$. For convenience, the dimensionless time $t_D$ is represented by $t$ itself. During the evolution of a tumour, significant changes occur in the tissue microenvironments, to which the tumour cells are exposed, so that the parameter $\lambda$, representing the coupling of the system to the environment, changes progressively as the tumour evolves and is also subjected to fluctuations. The average value of $\lambda$ changes from positive to negative as the growth rate constant $k_2$, associated with the CP population, exceeds $k_1$, the growth rate constant of the NCP subpopulation. A positive (negative) value of $\lambda$ favours the NCP (CP) subpopulation with fractional value $q$ $((1-q))$. The fluctuations in the dynamics are taken into account by replacing $\lambda$ by $\lambda + \sigma\xi$. Let $p(q,t)dq$ denote the probability of finding $q$ in the interval $(q, q+dq)$ at time $t$. In the Ito formalism, the stochastic differential rate equation (7) yields the Fokker-Planck equation (FPE):

$$\frac{\partial}{\partial t}p(q,t) = -\partial_q[f(q)p(q,t)] + \frac{\sigma^2}{2}\partial_{qq}[g^2(q)p(q,t)] \qquad (8)$$

where $f(q) = \frac{1}{2} - q + \lambda q(q-1)$. Note that the noise has the character of a multiplicative noise as it depends on the state of the system. In the case of additive noise, there is no state dependence and the FPE has the form as in (8) with $g(q) = 1$. From (8), the steady state probability density function (PDF) is given by

$$p_S(q) = \frac{K}{[q(1-q)]^2} \exp\left[-\frac{2}{\sigma^2}\left(\frac{1}{2q(1-q)} + \lambda \ln\left(\frac{1-q}{q}\right)\right)\right] \qquad (9)$$

where $K$ represents the normalization constant. The extrema, $q_m$, of the PDF $\left(\frac{dp_S(q)}{dq} = 0\right)$ are obtained from the solution of the equation [7]

$$\frac{1}{2} - q_m + \lambda q_m(1-q_m) - \sigma^2 q_m(1-q_m)(1-2q_m) = 0 \qquad (10)$$

The significance of the extrema lies in the fact that they are the suitable indicators of a transition [7]. In a unimodal to a bimodal distribution, the number of maxima of $p_S(q)$ change from one to two. The maxima have a natural interpretation in terms of the 'phases' of the system since they denote the most probable values preferentially observed in experiments. Knowing $p_S(q)$, one can define a stochastic potential, $\varphi(q)$ as

$$p_S(q) = K \exp\left[-\frac{2}{\sigma^2}\varphi(q)\right] \quad (11)$$

From (9), the explicit expression for $\varphi(q)$ is

$$\varphi(q) = \frac{1}{2q(1-q)} + \sigma^2 \ln[q(1-q)] + \lambda \ln\frac{1-q}{q} \quad (12)$$

The maxima of $p_S(q)$ correspond to the valleys of the potential whereas a minimum becomes a hilltop. In the case deterministic dynamics, the deterministic potential, $V(x)$, is defined as $\frac{dx}{dt} = f(x) = -\frac{dV}{dx}$. The minima of $V(x)$ are the stable steady states of the dynamics. In the case of additive noise, the deterministic and stochastic potentials are identical modulo an inessential constant so that the minima of the stochastic potential are the macroscopic steady states. This is also true in the case of the multiplicative noise if the noise intensity $\sigma^2$ is low. If the noise intensity is sufficiently large, the stochastic potential may develop new features like the appearance of new extrema. This is clear from (10) in which the first two terms contribute to the deterministic steady state equation and the last term represents the effect of the external noise. In the case of additive noise, the last term does not contribute as $g(q) = 1$ and the extrema of the steady state PDF coincide with the deterministic steady states (in the case of the tumour model there is one unique steady state, shown in (2), with β replaced by λ). In the ball-in-a landscape picture, the effect of the additive noise is to jiggle the ball around in the potential landscape with the potental retaining its shape. The valleys and the hilltops in the deterministic case retain their identity in the presence of aditive noise. In the case of multipicative noise and low noise intensity, the number and the nature of the steady states do not change from those in the deterministic case. For large noise intensity, the possibility occurs that the solutions of (10) do not match in number and position with those in the deterministic case. Now the ball not only jiggles around in the landscape but the shape of the potential also changes from, say, a single-valley structure to one with two valleys.

Returning to (10), the roots of the equation, defining the extrema of $p_S(q)$ are for $\lambda = 0$:

$$q_{m0} = \frac{1}{2}, q_{m\pm} = \frac{1}{2}\left[1 \pm \left(1 - \frac{2}{\sigma^2}\right)^{\frac{1}{2}}\right] \quad (13)$$

From (13), it is clear that a purely noise-induced transition occurs at a critical value $\sigma_c^2 = 2$ of the noise-intensity from a unimodal (a single maximum $q_{m0}$) to a bimodal (two maxima $q_{m+}$, $q_{m-}$ and a minimum $q_{m0}$) steady state PDF $p_S(q)$. Figure 1 shows the plots of $p_S(q)$ versus $q$ for $\sigma^2 < \sigma_c^2, \sigma^2 = \sigma_c^2$ and $\sigma^2 > \sigma_c^2$. The PDF is flat-topped at the critical noise-intensity, characteristic of a critical-point transition.

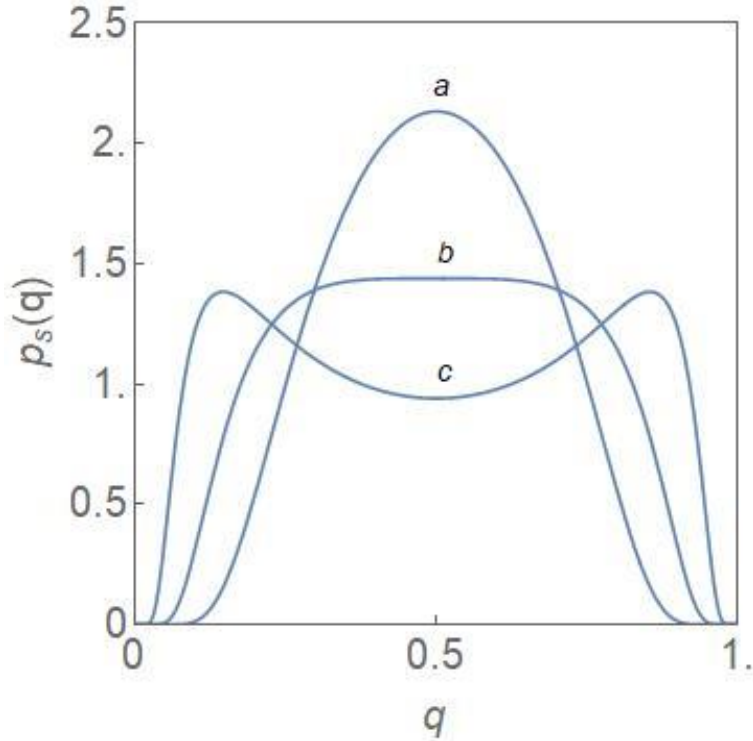

Figure 1. Steady state PDF, $p_S(q)$, versus $q$ for $\lambda = 0$. The noise intensity $\sigma^2$ has values 1.0 ($a$), 2.0 ($b$) and 4.0 ($c$). A purely noise-induced transition from unimodality to bimodality occurs when the noise intensity exceeds the critical value $\sigma_c^2 = 2$.

1.3 Critical Exponents

The critical exponents for the purely-noise-induced transition in the population genetics model are calculated using some standard procedure [7]. For biochemical models with positive feedback and bistable deterministic dynamics, Bose and Ghosh [10] have shown that the deterministic steady state equation serves as a starting point for the calculation of the critical exponents if the noise is additive in character. The method is generalised to the case of multiplicative noise by starting with (10), the solutions of which are the extrema of the steady state PDF. The proposed method is applicable irrespective of the character of the noise, additive, multiplicative or with both the components present. We rewrite (10) as

$$F(q_m) = \frac{1}{2} - q_m + \lambda q_m(1-q_m) - \sigma^2 q_m(1-q_m)(1-2q_m) = 0 \quad (14)$$

In the absence of noise, the equation reduces to the deterministic steady state equation. We briefly sketch the method of derivation of the critical exponents [10]. The idea is to reduce the equation $F(q_m) = 0$ to the form of the equation of state of the mean-field Ising model close to its critical point expressed as [24]

$$h - \theta m - \frac{1}{3}m^3 = 0 \quad (15)$$

In (15), $m$ represents the average magnetization per spin, $\theta = \frac{(T-T_c)}{T_c}$ is the reduced temperature with $T_c$ being the critical temperature and $h$ is the reduced magnetic field. The order parameter of the transition is the spontaneous magnetization m ($h = 0$) which has a zero value for $T > T_c$ and a non-zero value for $T < T_c$. The critical point is given by $\theta = 0, h = 0$. The equation of state further corresponds to the normal form of the steady state equation of an imperfect SP bifurcation [25]:

$$ry - y^3 + H = 0 \quad (16)$$

with the bifurcation point, $r = 0, H = 0$, separating a regime of monostability from that of bistability.

To proceed with the derivation, the expression $F(q_m) = 0$ is Taylor expanded around a point $q_c$ to the third order with the proviso that $F''(q_c) = 0$, so that a second-order term is absent in the Taylor expansion as in the case of the magnetic equation of state (15). The expansion has the form

$$F(q_c) + F'(q_c)(q_m - q_c) + F'''(q_c)\frac{((q_m - q_c))^3}{3!} = 0 \quad (17)$$

Comparing with (15), one can write down the following relations between the thermodynamic and dynamic quantities:

$$m = \frac{((q_m - q_c))}{q_c}, h = -\frac{2F(q_c)}{F'''(q_c)q_c^3}, \theta = \frac{2F'(q_c)}{F'''(q_c)q_c^2} \quad (18)$$

For the model under consideration,

$$q_c = \frac{1}{2} - \frac{\lambda}{6\sigma^2} \quad (19)$$

At the critical point, the order parameter $m = 0$, i.e., $q_C = q_m$. Also, $\lambda = 0$ and $\sigma^2 = 2$. Thus, (10) has a triple root $q_m = q_{m0} = q_{m\pm} = q_C = \frac{1}{2}$. In terms of the dynamical quantities, one gets $F(q_C) = 0, F'(q_C) = 0$ at the critical point, i.e., the SP bifurcation point. One can check that $F'(q_C)$ changes sign at the critical noise intensity $\sigma_C^2 = 2$ and $F'''(q_C)$ is always negative. These features are consistent with the change in sign of the magnetic quantity, in (18), at the critical temperature. The noise intensity plays the role of temperature in the nonequilibrium model. The critical behaviour is identical to that of the mean-field Ising model with the parameter $\lambda$ playing the role of the magnetic field. The order parameter $m = q_{m+} - 0.5$ exhibits power-law singularity close to the critical point, $m \sim (\sigma^2 - \sigma_C^2)^{1/2}$ with the critical exponent having the value $\frac{1}{2}$. Also, $m \sim \lambda^{1/3}$ at the critical point, $\sigma^2 = \sigma_C^2 = 2$ with the critical exponent $\delta = 3$ and $\frac{\partial m}{\partial \lambda}|_{\lambda=0} \sim |\sigma^2 - \sigma_C^2|^{-1}$ with the criticaL exponents $\gamma = \gamma' = 1$.

2. Nonequilibrium Model with $\lambda \neq 0$

The critical point transition, belonging to the mean-field Ising universality class, requires the parameter $\lambda$ to be zero. When $\lambda$ has a finite value, the transition from unimodality to bimodality occurs when the noise-intensity exceeds a critical value dependent on the value of $\lambda$. The bifurcation curves are obtained in a parametric form since the parameter $\lambda$ cannot be written explicitly as a function of $\sigma^2$. The region of bistability is enclosed within a pair of bifurcation boundaries with each boundary representing a line of SN bifurcations, equivalent to first-order phase transitions [25]. The computation of the boundaries is carried out in the following manner. We start with (14) and note that for a SN bifurcation to occur, the conditions to be satisfied are $F(q_m) = 0$ and $F'(q_m) = 0$. The two conditions yield the expressions

$$\sigma^2 = \frac{1}{2}\frac{1}{q_m(1-q_m)} + \frac{\lambda}{(1-2q_m)} \qquad (20)$$

$$\sigma^2 = \frac{-1 + \lambda(1-2q_m)}{1 - 6q_m + 6q_m^2} \qquad (21)$$

From (20) and (21), one obtains the parametric relations

$$\lambda = \frac{(1-2q_m)^3}{4q_m^2(1-q_m)^2} \qquad (22)$$

$$\sigma^2 = \frac{1}{4(1-q_m)^2} + \frac{1}{4q_m^2} \qquad (23)$$

The bifurcation curves are defined by (22) and (23). For $0 < q_m < 1$, the points $(\sigma^2(q_m), \lambda(q_m))$ are plotted in the $(\sigma^2 - \lambda)$ plane (Figure 2). Note that the variable $q_m$ represents an extremum (maximum) of the steady state PDF so that the phase diagram has a stochastic character.

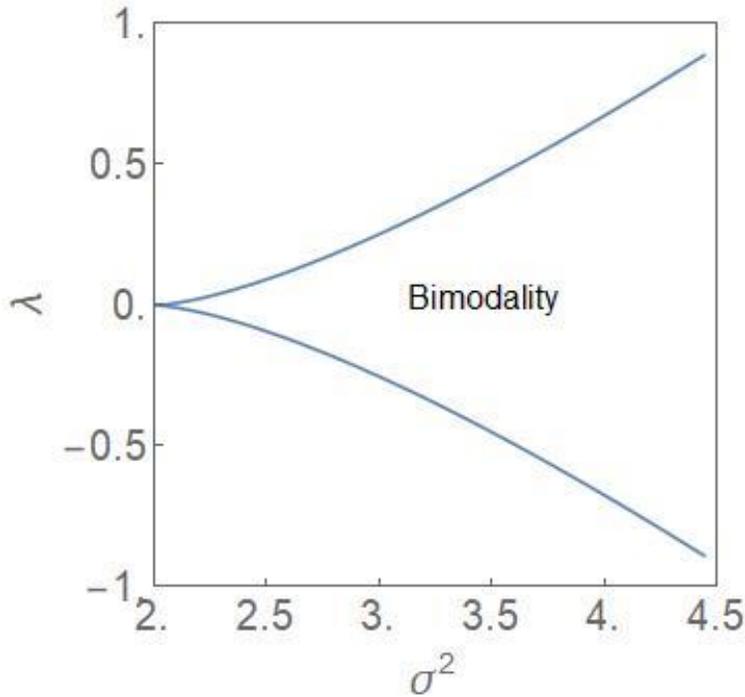

Figure 2. Parametric plots of bifurcation curves (22) and (23) in the $(\lambda - \sigma^2)$ plane. The curves separate a region of bimodality from the regions of unimodality.

One identifies a region of bimodality separating two regions of unimodality. In the bimodal (unimodal) region, the steady state PDF, $p_S(q)$, has two peaks (one peak). The critical point is depicted by the point $(2,0)$. The termination of the bifurcation curves at this point is reminiscent of a line of first-order transitions terminating at a critical point in the cases of equilibrium thermodynamic phases transitions like the liquid-gas and the paramagnetic-ferromagnetic phase transitions [24]. In the deterministic case, the region of bimodality disappears and one has only monostability in the full parameter regime. In the stochastic case, the bimodality is accompanied by hysteresis [6,7] the plots of which are computed from (14) for both $q_m$ and $(1 - q_m)$, the population fractions, corresponding to the two subpopulations. Figure 3 shows the hysteresis curves ($\sigma^2 = 4$) and one finds a reflection symmetry, between the curves, across the vertical line at $\lambda = 0$.

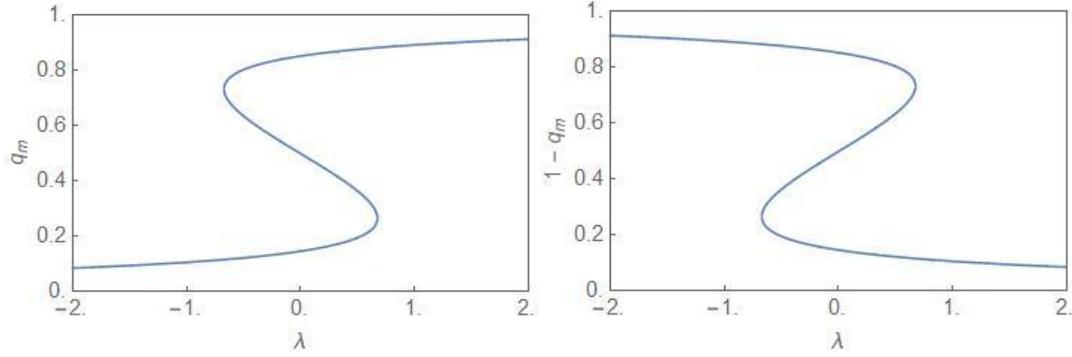

Figure 3. Hysteresis plots for $q_m$ and $1 - q_m$ versus $\lambda$ with $\sigma^2 = 4$. The two plots correspond to the two phenotypically distinct subpopulations.

3.1 Cancer as a Phase Transition

The heterogeneity in the tumour population is described in terms of the two subpopulations: NCP and CP. Examining the expression for the steady state PDF, $p_S(q)$, in (9), one finds that for $\lambda = 0$ the PDF has exchange symmetry. It remains invariant under the transformation $q \to 1 - q$, i.e., when the subpopulation fractions are interchanged. When $\lambda$ has finite values, the exchange symmetry is broken and one finds the relation

$$P_S(q, \lambda) = P_S(1 - q, -\lambda) \qquad (24)$$

where $q$ and $1 - q$ represent the fractions of the NCP and CP subpopulations, respectively, of the total population. We first consider the case $\lambda = 0$. This is the situation when, on an average, neither of the subpopulations is favoured by the environment. When the noise-intensity $\sigma^2$ is less than the critical value $\sigma_c^2 = 2$, the steady state PDF, $p_S(q)$, has a unique maximum at $q_{m0} = \frac{1}{2}$ which coincides with the deterministic stable steady state. When the noise intensity exceeds the critical value, $p_S(q)$ becomes bimodal with the peak positions at $q_{m\pm}$ (see (13)). One notes the relation $q_{m+} = 1 - q_{m-}$ with $q_{m-} < q_{m+}$. For each subpopulation, the fractional value $q$ has an equal probability (Figure 1) to peak at a low or a high value. Since $P_S(q) = P_S(1 - q)$, if $q$ for the NCP subpopulation is low (high), that of the CP subpopulation is high (low). One designates the $\lambda = 0$ state as a state of balance between the subpopulations since each of them has equal probability to have a larger fractional value. The peak positions $q_{m-}$ and $q_{m+}$ move respectively to 0 and 1 respectively as the noise intensity $\sigma^2 \to \infty$.

A simple analogy with a population of colour-coded chemical reactants provides a clear physical picture of the population heterogeneity in the two distinct regimes: $\sigma^2 < \sigma_c^2$ and $\sigma^2 > \sigma_c^2$ for $\lambda = 0$ [7]. The dynamics of the system of reactants are as given in (1). There are two subpopulations of reactants of yellow and blue colour with the fractional values $x$ and $1 - x$ respectively. For $\sigma^2 < \sigma_c^2$, neither subpopulation dominates and the reacting system exhibits a flickering green colour. Above the critical noise-intensity, the reacting system will either be predominantly yellow or predominantly blue with equal times, on an average, spent in both the states. In terms of the cancer model, the population is mostly NCP or CP, each possibility occurring with equal probability. When $\lambda$ has a finite value, the balance tilts in favour of the NCP (CP) subpopulation for $\lambda > 0$ ($\lambda < 0$). Figure 4 shows the cumulative distribution function (CDF) of the random variable $q$ versus $\lambda$ with the CDF defined as

$$F_S(q^*) = \int_0^{q^*} p_S(q)\, dq \quad (25)$$

$F_S(q^*)$ gives the probability that the fractional value $q$ of the NCP subpopulation is less than or equal to $q^*$. The curves labeled "d" and "e" correspond to $F_S(0.5)$ and $1 - F_S(0.5)$ respectively.

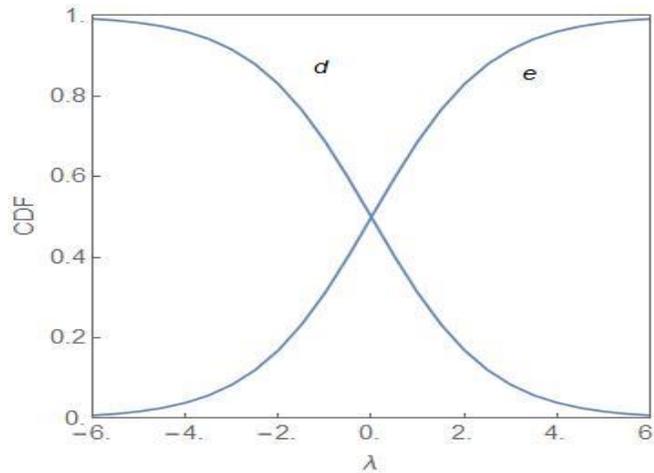

Figure 4. Cumulative distribution function (CDF) of $q$ versus $\lambda$ for the cases $(d)\ F_S(0.5)$ and $(e)\ 1 - F_S(0.5)$ with $\sigma^2 = 8.0$.

From Figure 4, it is clear that the NCP subpopulation becomes the dominant one as λ becomes positive (Curve *e* lies above Curve *d*), the dominance becoming more prominent as λ increases towards more positive values. For a sufficiently large value of λ, the CDF reaches its maximal value of 1. For the CP subpopulation, the plots $d$ and $e$ represent the CDFs $1 - F_S(0.5)$ and $F_S(0.5)$ respectively. The subpopulation becomes more and more dominant as $\lambda$ acquires more negative values. Figure 5 shows the evolution of $P_S(q)$ as $\lambda$ changes from positive to negative values and with $\sigma^2 = 4.0$. Invoking the relation in (24), the same set of plots represent the evolution of $P_S(1-q)$ but with the sign of $\lambda$ changed. In the region of bimodality and for $\lambda = 0$, the NCP and CP subpopulations are in a balanced state. The probability that the NCP (CP) subpopulation becomes more dominant is enhanced as $\lambda$ increases towards positive (negative) values.

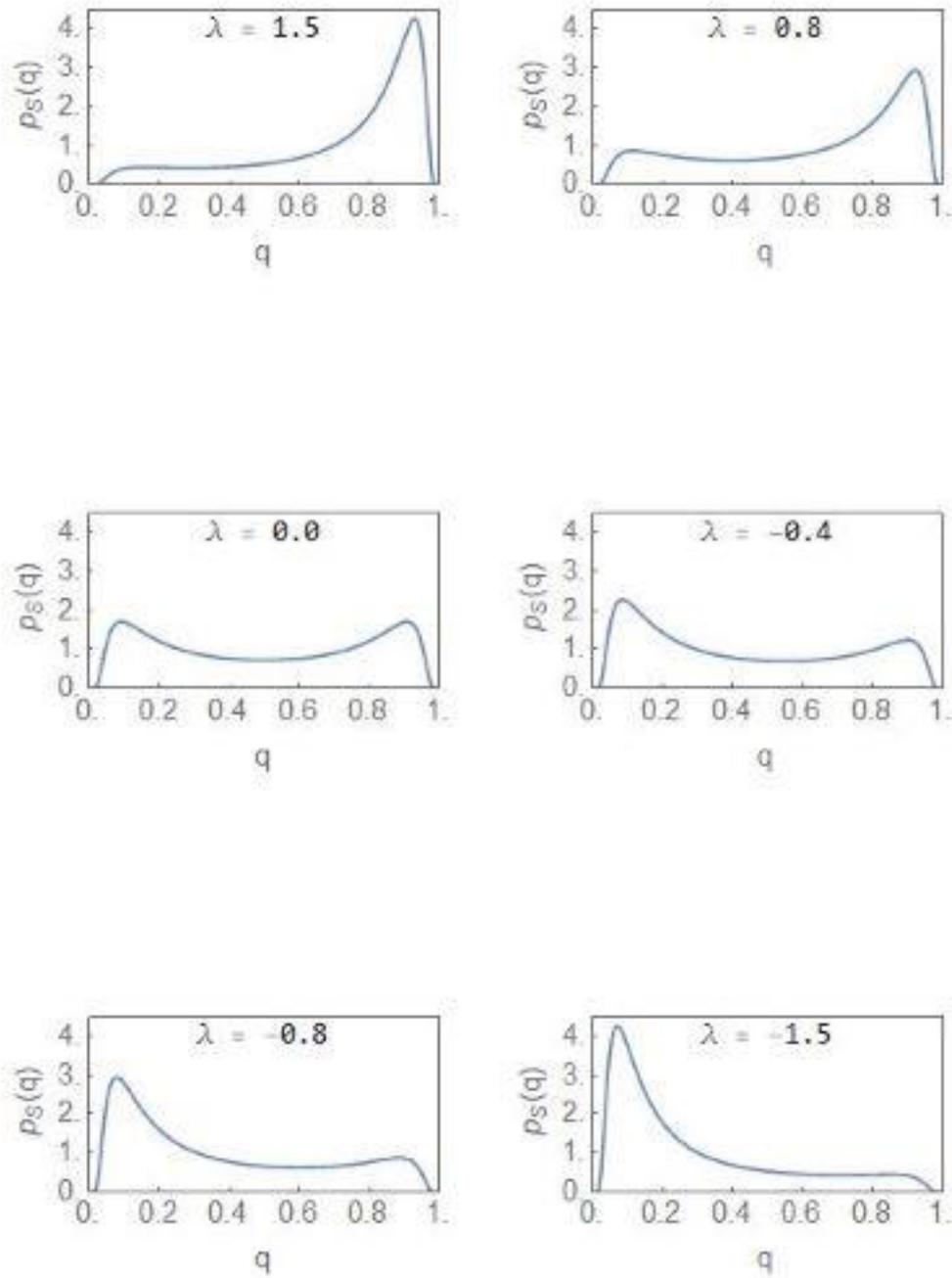

Figure 5. Evolution of the steady state PDF $P_S(q)$ versus $q$ as the parameter decreases from positive to negative values with $\sigma^2 = 4.0$. The same set of plots represent the evolution of $P_S(1-q)$ versus $q$ but with the sign of $\lambda$ changed.

In the regions of unimodality ( $\lambda = 1.5, \lambda = -1.5$), the steady state PDF has a single peak with the NCP (CP) subpopulation dominating over the other subpopulation for positive (negative) λ value. Within the region of bimodality, the opportunity exists to reverse the dominance of the CP population through appropriate strategies including drug therapy but once the bifurcation boundary (Figure 2) is crossed into the region of unimodality, irreversibility in the form of hysteresis (Figure 3) sets in. The sudden deterioration from the pre-disease to the disease state at the bifurcation point effectively constitutes a 'point of no return'. Figure 6 shows the evolution of the stochastic potential $\varphi(q)$ as a function of λ in a few representative cases with $\sigma^2 = 4.0$.

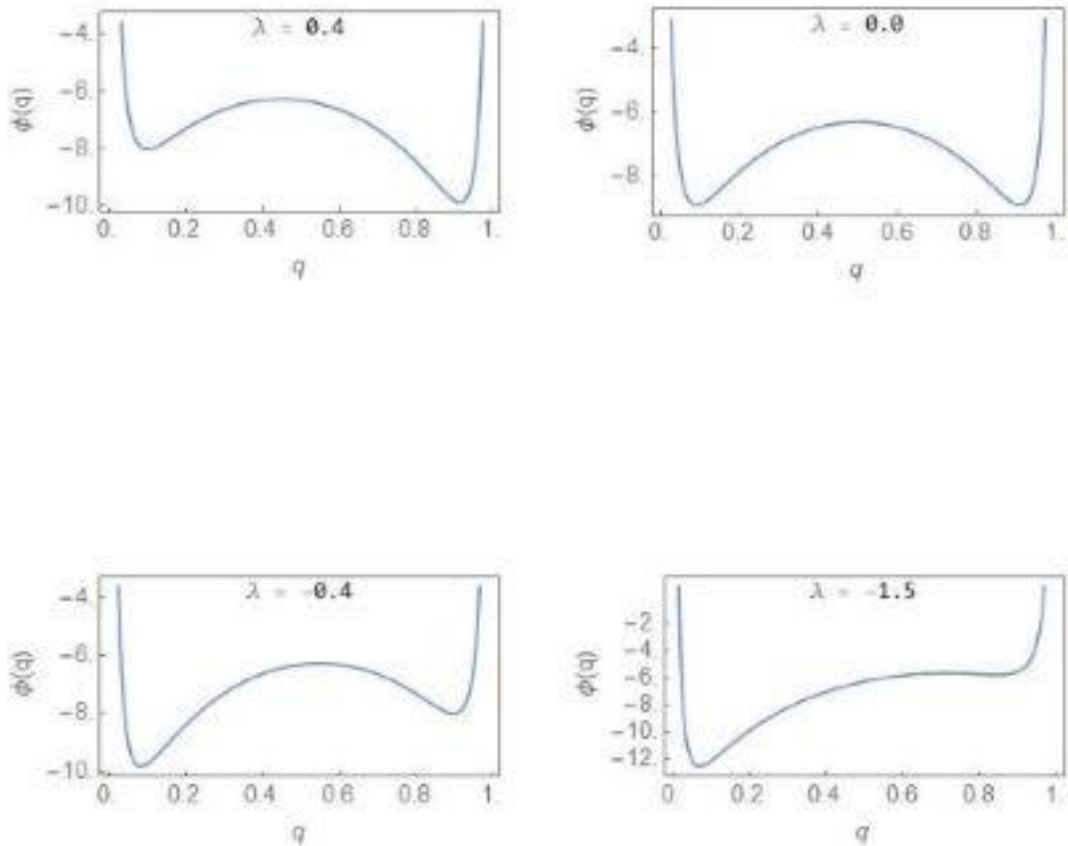

Figure 6. Stochastic potential $\varphi(q)$ ((12)) versus $q$ with $\sigma^2 = 4.0$.

From Figure 6, one finds that in the case of $\lambda = -1.5$, a very small fraction of the total population belongs to the NCP subpopulation so that the CP subpopulation constitutes the major component of the tumour population. The relative stability between the NCP and CP subpopulations shifts with changes in the parameter λ.

The key point emerging from our study is that the progression of a tumour population from the normal to the cancer state can be understood in terms of the critical-point and first-order phase transitions [26, 27,28] which occur in the nonequlibrium (dynamical phase transitions). In the normal state, the NCP and the CP subpopulations are kept in balance with both the subpopulations equally favourable. This is specially so for $\lambda = 0$ and the noise-intensity $\sigma^2 < \sigma_c^2$. A critical-point transition to a region of bimodality occurs at the critical noise-intensity (Figure 1) with the steady state PDF developing two peaks around the states, $x_{m-}$ and $x_{m+} = 1 - x_{m-}$. The fractional values, $x_{m-}$ and $x_{m+}$ occur with equal probability so that the subpopulations continue to remain in balance. The critical-point transition into the region of bimodality opens up the possibility of one subpopulation outcompeting the other. This asymmetric situation or tilting of the balance is possible when the parameter $\lambda$ has a non-zero value with the NCP (CP) subpopulation becoming dominant for positive (negative) values of λ. When the growth rate constant of the CP subpopulation exceeds that of the NCP subpopulation, the parameter λ becomes negative and the pre-disease state is reached at the bifurcation boundary in Figure 2. Up to the pre-disease state, the deterioration towards the disease state is reversible and the opportunity for damage control exists. In the region of unimodality and for negative values of λ, the disease state is reached with the CP subpopulation achieving irreversible dominance. This is clear from the hysteresis curves ( Figure 3) which show that to return to the original branch, once a bifurcation has taken place, an "overshoot" of the control parameter value is requred. In the region of bimodality, the upward and downward jumps occur at the same parameter values, i.e., the transitions are reversible.

3. Quantitative Signatures of the Onset of Dominance

Cell-fate transitions, as in the cases of cell differentiation and the sudden deterioration in the progression of complex diseases like cancer, share similarities with phase transitions [17, 26, 29, 30]. Rapid advances in single-cell techniques have made it possible to obtain network-wide gene expression data at single-cell resolution for a large ensemble of cells. The large-scale data provide signatures of dynamical bifurcations like the pitchfork and SN bifurcations, coinciding with experimentally observed cell-fate transitions. The pre-disease state in the route from the normal to the disease state marks the onset of irreversibility as in the case of the SN bifurcation. In literature the transition to irreversibility is referred to as the tipping point and sometimes as the "critical-point" transition [20] though the latter nomenclature is reserved in the physics literature for continuous transitions. The method of dynamical network biomarker (DNB) has been developed to detect the early warning signal (EWS) of the approach to the pre-disease state, in the proximity of a bifurcation point [17, 18]. The DNB consists of a dominant group of genes in the gene regulatory network which provides the EWS in terms of the following features of the experimental data: (i) the coefficient of variation (CV) of the expression of a DNB gene increases significantly as the bifurcation point is approached, (ii) an increased covariation between the members of the DNB and (iii) the rapid decrease in the correlation between a DNB and a non-DNB element of the network. Appropriately constructed criticality indices capture the increased covariation in the vicinity of the tipping point [17, 18]. The DNB method has been applied successfully to high-throughput experimental data towards the early detection of the pre-disease state in a number of complex diseases including cancer.

In recent studies, some specific entropic measures have been proposed to detect the pre-disease state from the data on reference samples and a single case sample. The concept of entropy is fundamental in statistical mechanics and information theory [31,32]. The entropy of a random variable provides a quantitative measure of the uncertainty associated with the random variable. On an average, it is a measure of the amount of information required for a description of the random variable. Various entropic measures have been proposed so far to analyse single-cell datasets obtained in high-throughput experiments. One of these measures is that of relative entropy, a measure of the distance between two probability distributions. In the case of discrete random variables, the relative entropy, in terms of the Kullback-Leibler (KL) distance [31,32], is defined as

$$D_{KL}(P||Q) = \sum_{x} P(x) \log \frac{P(x)}{Q(x)} \quad (26)$$

where $P(x)$ and $Q(x)$ represent the probability mass functions. We take the base of the logarthm to be 2. In the case of continuous random variables, the summation is replaced by an integral with $P(x), Q(x)$ representing the PDFs. The KL distance provides a measure of the loss in information when an approximation $Q(x)$ is used for the true mass function $P(x)$. The KL distance is always positive and has the value zero only if $P(x) = Q(x)$. It is, however, not a bonafide distance measure as it is not symmetric and does not satisfy the triangle inequality. The Jensen-Shannon divergence is defined to be [33,34]

$$D_{JS}(P||Q) = \frac{1}{2} D_{KL}(P||M) + \frac{1}{2} D_{KL}(Q||M), M = \frac{1}{2}(P + Q) \quad (27)$$

One can check that $D_{JS}(P||Q) = D_{JS}(Q||P)$ (symmetry) and $0 \leq D_{JS}(P||Q) \leq 1$. The zero (one) value indicates perfect (zero) overlap between the distributions. Also, the quantity $\sqrt{D_{JS}(P||Q)}$ defines a metric distance satisfying the triangular inequality. The definitions (26) and (27) hold true when $P$ and $Q$ represent the cumulative probability. Based on the last definition, incorporated in a single-sample-based Jensen-Shannon Divergence (sJSD) method, a criticality index ICI (inconsistency index) has been computed from the real datasets of a few complex diseases including some specific types of cancer [34]. The criticality index exhibits a sharp peak as a signature of a tipping point transition from a pre-disease to a disease state.

In our two-subpopulation model of tumour evolution, the key dynamical variables are the fractions $q$ and $(1 - q)$ of the NCP and CP subpopulations, respectively, in the total population. In this case, as already discussed, the NCP (CP) subpopulation becomes progressively more dominant as the value of λ increases towards more positive (negative) values. In the following, taking cues from the EWS of the approach to the pre-disease state [17, 18], we provide three quantitative signatures of the transition from the state of balanced subpopulations, $\lambda = 0$, to the emergence of a dominant subpopulation. We confine our attention to the parameter regime, $\sigma^2 > \sigma_C^2$. Figure 7 shows the plot of the variance of the steady state probability distribution ((9)) versus the parameter $\lambda$ for $\sigma^2 = 4$. The approach to the state of balance from both the sides is indicated by a rise in the magnitude of the variance with the peak value occurring in the state of balance itself. If initially the NCP subpopulation is dominant, an increase in the variance would indicate the approach to the state of balance. The state of balance effectively serves as a tippng point from the dominance of the NCP subpopulation to that of the CP subpopulation. The deterioration can be reversed as long as the system is in the region of bimodality with irreversibility setting in as the bifurcation boundary is crossed.

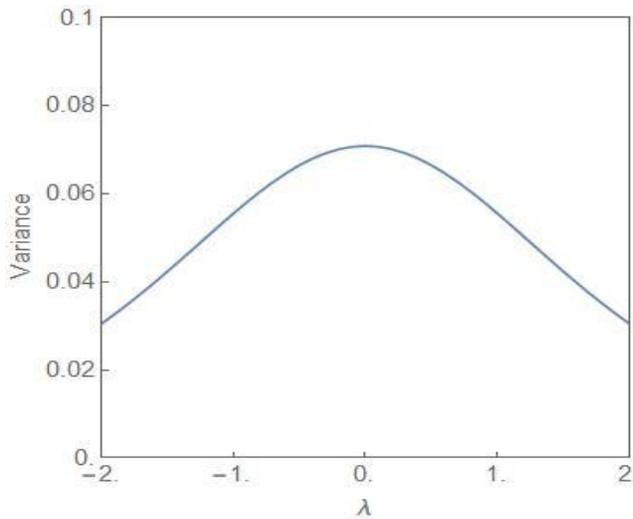

Figure 7. The variance of the steady state probability distribution, given in (9), versus λ for $\sigma^2 = 4$.

Figure 8 shows the plot of the Jensen-Shannon divergence, $D_{JS}(P||Q)$ ((27)) [34], versus the parameter λ. The steady state probability distributions $P$ and $Q$ are as given in (9) with $\sigma^2 = 4$, $\lambda \neq 0$ for the distribution $P$ and $\sigma^2 = 2, \lambda = 0$ for the distribution Q. The latter distribution is the flat-topped critical probability distribution shown in Figure 1. Thus the JS divergence provides a measure of how different the distribution $P$ is from the critical distribution as the parameter λ is varied. From the Figure it is clear that the closest similarity is for λ = 0. Away from this value, the difference between the two distributions become more and more dissimilar as λ increases towards more positive/negative values. Figure 9 shows the plot of $D_{JS}(P||Q)$ versus the parameter λ but now $\sigma^2 = 4$ for both the distributions $P$ and $Q$ but the λ-values have opposite signs. Since the NCP and CP subpopulations become dominant for the positive and negative values of λ respectively, the JS divergence captures the dissimilarity of the two situations, one favouring the NCP and the other favouring the CP subpopulation.

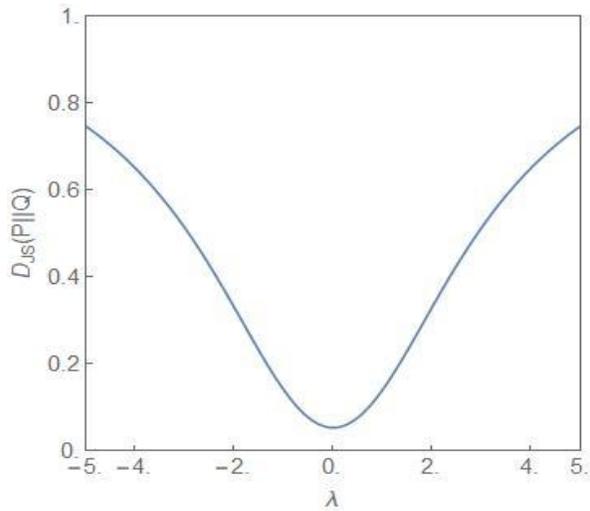

Figure 8. The plot of $D_{JS}(P||Q)$ versus λ. For the P-distribution, $\sigma^2 = 4$, $\lambda \neq 0$. For the Q-distribution, $\sigma^2 = 2, \lambda = 0$, i.e., the distribution is the critical distribution.

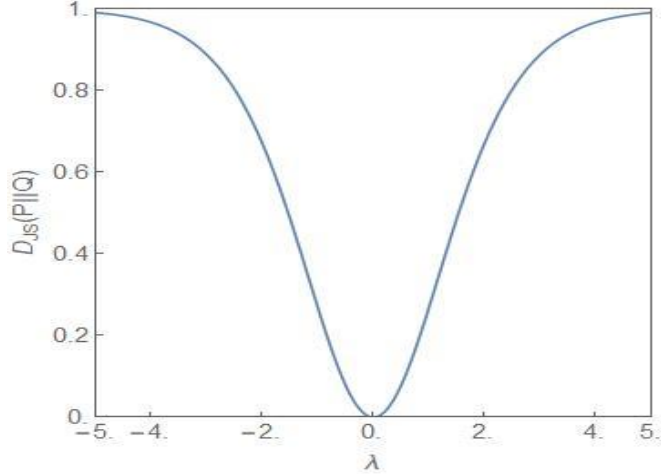

Figure 9. The plot of $D_{JS}(P||Q)$ versus $\lambda$. For both the distributions $P$ and $Q$, $\sigma^2 = 4$ but the $\lambda$ values are opposite in sign.

The two distributions are identical when $\lambda = 0$ and have no overlaps when $D_{JS}(P||Q) = 1$. The third quantitative signature is provided by the entropic measure cumulative paired entropy (*CPE*) defined as [35]

$$CPE = -\int F(x)\log F(x)dx - \int (1 - F(x))\log(1 - F(x))\,dx \quad (28)$$

where $F(x)$ is the CDF of the random variable $X$. If $X$ has a compact support $[a, b]$ then the *CPE*, without any constraints, attains its maximum value for $F(x) = \frac{1}{2}$ and $a \leq x < b$. The associated distribution is designated as the bipolar distribution with $P(X = a) = P(X = b) = \frac{1}{2}$. Figure 10 shows the plot of the *CPE* versus the parameter $\lambda$ for $\sigma^2 = 4.0$. The plot is symmetric around $\lambda = 0$ with only the positive half shown in the Figure.

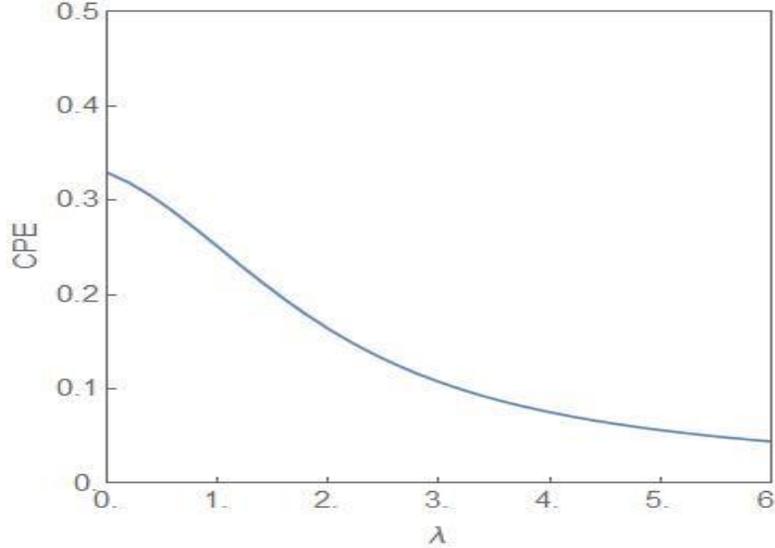

Figure 10. The plot of $CPE$ ((28)) versus λ with $\sigma^2 = 4.0$. The entropic measure has the highest value at λ = 0.

The bipolar distribution which maximizes the $CPE$, in the case of the two-subpopulation tumour model, is obtained in the limit of $\sigma^2 \to \infty$ and $\lambda = 0$, with $p_S(q=0) = p_S(q=1) = \frac{1}{2}$ and $[a,b] = [0,1]$ [7]. In the bipolar limit, the $CPE$ has the value 1. The physical interpretation is that with equal probability $\left(=\frac{1}{2}\right)$, all the cells in the tumour population belong to either the NCP or the CP subpopulation. The situation is characterized by the term "contradictory information (CI)" [35] in contrast to the case of "minimum information", a feature of the uniform distribution. In the latter case, the random variable $X$ has an equal probability to assume any value in a range of values whereas in the former case the variable has a choice between two extreme values with equal probability. Another interesting feature of the bipolar limit is that the transitions between the peaks at $q = 0$ and 1 are extremely rare so that the stabilization of one phenotype occurs with respect to the other phenotype [7]. When $\lambda \neq 0$, the growth rates of the subpopulations become unequal but the so-called "unfit" phenotype may still be stabilized. This happens because even if the system is not at the bipolar limit, the transitions between the probability peaks are still infrequent.

4. Conclusions

Population heterogeneity, in terms of distinct phenotypes, is a widely observed feature in cellular populations. The heterogeneity is mostly driven by noise, of both intrinsic and extrinsic origins, and has a significant role in cellular decision-making processes as well as in the implementation of the bet-hedging strategies for the survival of populations [1, 2, 3]. In the presence of noise, the dynamics have a stochastic component and the attractors of the dynamics in the state space describe the "stable" states of the system. An attractor corresponds to a cloud of points in the state space which reduces to a single point or a cycle in the limit of deterministic dynamics. The cell population dynamics are subjected to two effectively opposing influences, one attracting the dynamics towards the cloud centre or to a cycle of states ( the "drift" term appearing as the first term on the right-hand-side of the FPE (8)) and the other giving rise to diffusion in the state space ( the "diffusion" or "noise" term appearing as the second term on the right-hand-side of the FPE). The idea of attractors as different cell types was demonstrated in a pioneering study by Kauffman [36] and forms the central concept in the analysis of experimental observation-based data [30, 37, 38]. In the context of a tumour population, Kauffman first proposed the hypothesis that the cancerous state could be an attractor of the tumour dynamics [39].

It is now well-recognized that "dramatic" changes in the tissue microenvironments are associated with the progression of tumours [12, 13] and these microenvironmental changes play an active role in tumorigenesis, specifically, in promoting the dominance of the CP subpopulation. One principal result of our study is to establish the equivalence between the two-subpopulation dynamical model of tumour progression, as described in (3) and (4), and the well-known population genetics model with dynamics as described in (1). The mapping enables one to reduce the number of parameters from four $(k_1, k_2, r_1, r_2)$ to one, namely, λ. The parameter λ takes care of the coupling to the environment and is expected to undergo large changes in its average value as a tumour progresses to the cancerous state. As shown in (6), the steady state subpopulation fractions, $(f_i, i = 1,2)$, are governed by the single parameter λ, which is a ratio of the parameters $k_{diff}$ and $2r_e$. The parameter is further subjected to considerable fluctuations due to the random variations in the tissue microenvironments. Because of the equivalence, some well-known results of the stochastic population genetics model are of validity in the case of the two-subpopulation tumour model. The most important of these is the existence of a noise-induced transition from unimodality to bimodality when the noise intensity exceeds a critical value. The maxima of the steady state PDF, which correspond to the most probable states, define the "phases" of the system. We have developed a procedure, based on the knowledge of the extrema of the steady state PDF((14)), to explore the phase diagram of the NCP-CP model exhibiting both critical-point and first-order phase transitions in the nonequilibrium. The critical exponents, associated with the critical-point transition $\lambda = 0, \sigma^2 = 2$, belong to the mean-field Ising universality class. The evolution of the steady state PDF and the stochastic potential as the parameter λ changes (Figures 5 and 6) capture the progression of the tumour population from a "healthy" to a "disease" state. In the former (latter) state, the NCP (CP) subpopulation becomes dominant. Some of the concepts and methodologies of the statistical mechanical description of phase transitions, like the state transition from a unimodal to a bimodal frequency distribution through a flattened unimodal profile (Figure 1) and the depiction of the state-transitions via potential landscapes and hysteresis curves (Figures 3 and 6), have made their appearance in recent studies on cancer. In such studies, cancer is interpreted in terms of a state-transition in the time series of transcriptome (gene expression) data with the transition points (bifurcation/phase transition points) identified as the points in state space at which dynamical regime shifts take place [17, 28, 34, 40, 41] .

In the NCP-CP model, apart from the usual bifurcation transitions of the SP and SN type, there is another transition from a state of balance to a state of dominance in terms of the NCP and CP subpopulations. This transition can be detected through the changes in the variance of the steady state PDF and the entropic measures like the JS divergence and the CPE as the parameter $\lambda$ changes. The use of information-theoretic measures to detect the pre-disease state at which the irreversible transition to the disease state occurs is now a standard tool in the hands of data analysts. Some strategies for escaping the cancer attractor have been proposed by Huang and Kauffman [42]. A noise-based strategy of limited application till now arises from the experimental evidence of noise modulators, which are chemical compounds and biochemical complexes [43, 44]. These modulators (noise enhancers) do not change the mean expression level and by regulating the noise intensity can bring about an escape from a specific, say cancer, attractor. The stochastic NCP-CP model demonstrates the key role played by multiplicative noise in the progression of a tumour population to the cancerous state in which the CP subpopulation becomes dominant. The model, the key variables of which are the subpopulation fractions, is simple enough to be analytically tractable and may serve as a starting point for the characterization of the ITH taking into account the random variability of the tissue microenvironment. The assumptions underlyng the tumour model, with the defining equations as in (3) and (4), have considerable experimental support [14,15]. The proposal of a purely noise induced transition brought about by multiplicative noise, as demonstrated in the case of the population genetics model, is realizable in a number of experimental systems [7]. The link between the experimentally motivated two-subpopulation tumour model and the population genetics model lays open the possibility of designing experiments to probe how the tumour microenvironment noise tilts the phenotypic composition of the tumour in favour of the cancer-promoting subpopulation.

Acknowledgement. IB acknowledges the support of NASI, Allahabad, India under the Honorary Scientist Scheme.

Appendix A.

The solutions of (3) and (4) are given by

$$n_1(t) = C_1 e^{\lambda_1 t} \frac{r_2}{\lambda_1 - k_1} + C_2 e^{\lambda_2 t} \frac{r_2}{\lambda_2 - k_1}$$

$$n_2(t) = C_1 e^{\lambda_1 t} + C_2 e^{\lambda_2 t}$$

$$f_1(t) = \frac{n_1(t)}{n(t)} = \frac{C_1 e^{\lambda_1 t} \frac{r_2}{\lambda_1 - k_1} + C_2 e^{\lambda_2 t} \frac{r_2}{\lambda_2 - k_1}}{C_1 e^{\lambda_1 t} \left(1 + \frac{r_2}{\lambda_1 - k_1}\right) + C_2 e^{\lambda_2 t} \left(1 + \frac{r_2}{\lambda_2 - k_1}\right)}$$

$$\lambda_1, \lambda_2 = \frac{(k_1 + k_2) \pm \sqrt{\Delta^2 + 4 r_1 r_2}}{2}, \Delta = (k_1 - k_2)$$

Since $\lambda_1 > \lambda_2$, one has $e^{(\lambda_2 - \lambda_1)t} \to 0$ in the limit of large times, i.e., $t \to \infty$. The common time-dependent factor, $e^{\lambda_1 t}$, in the numerator and the denominator then cancels out resulting in the steady state expression for $f_1$ as

$$f_1 = \frac{r_2}{\lambda_1 - k_1 + r_2}$$

For the tumour model considered, $r_1 = r_2 = r_e$ and $f_1$ simplifies to

$$f_1 = \frac{1}{2\Delta} \left( (\Delta - 2 r_e) + \sqrt{\Delta^2 + 4 r_e^2} \right)$$

With the parameter $\lambda$ defined as $\lambda = \frac{\Delta}{2 r_e}$, the final expression for $f_1$ is

$$f_1 = \frac{1}{2\lambda} \left( \lambda - 1 + \sqrt{1 + \lambda^2} \right)$$

The steady state expression is identical to that for the genetic model, as given in (2), with the parameter β replaced by the parameter λ. The steady state in the tumour model, with equal asymmetric cell division rate constants, $r_1 = r_2 = r_e$, is governed by a single parameter $\lambda = \frac{k_1 - k_2}{2 r_e}$ rather than by the three parameters $k_1$, $k_2$ and $r_e$ independently.

References

1. Kærn, M.; Elston, T. C.; Blake, W. J.; Collins, J. J. Stochasticity in gene expression: from theories to phenotypes. *Nat. Rev. Genet*. **2005**, 6, 451-464.
2. Raj, A.; van Oudenaarden, A. Nature, nurture or chance: stochastic gene expression and its consequences. *Cell* **2018**, 135, 216-226.
3. Balázsi, G.; van Oudenaarden, A.; Collins, J. J. Cellular decision making and biological noise: from microbes to mammals. *Cell* **2011**, 144, 910-925.
4. Veening, J.-W.; Smits, W. K.; Kuipers, O. P. Bistability, epigenetics and bet-hedging in bacteria. *Annu. Rev. Microbiol* **2008**, 62, 193-210.
5. To, T. L.; Maheshri, N. Noise can induce bimodality in positive transcriptional feedback loops without bistability. *Science* **2010**, 327, 1142-1145.
6. Arnold, L; Horsthemke, W; Lefever, R. White and coloured external noise and transition phenomena in nonlinear systems. *Z. Physik B* **1978**, 29, 367-373.
7. Horsthemke, W; Lefever, R. *Noise-induced transitions*, 1st ed.; Springer-Verlag:Berlin,Germany,1984; pp.108-200.
8. Erez, A.; Byrd, T. A.; Vogel, R. M.; Altan-Bonnet, G.; Mugler, A. Universality of biochemical feedback and its application to immune cells. *Phys. Rev. E.* **2019**, 99, 022422.
9. Sornette, D. *Critical phenomena in natural sciences*, 2nd ed.; Springer-Verlag: Berlin, Germany, 2006; pp. 241-265.
10. Bose, I; Ghosh, S. Bifurcation and criticality. *J. Stat. Mech.: Theory and Experiment* **2019**, 043403.
11. Muñoz, M. A. Criticality and dynamical scaling in living systems. *Rev. Mod. Phys.* **2018**, 90, 031001.
12. Marusyk, A.; Almendro, V.; Polyak, K. Intra-tumour heterogeneity: a looking glass for cancer? *Nature Reviews Cancer* **2012**, 12, 323-334.
13. Marusyk, A.; Janiszewska, M.; Polyak, K. Intratumour heterogeneity: the Rosetta Stone of therapy resistance. *Cancer Cell* **2020**, 37, 471-484.
14. Jordan, N. V. et al. HER2 expression identifies dynamical functional states within circulating breast cancer cells. *Nature* **2016**, 537, 102-106.
15. Li, X.; Thirumalai, D. A mathematical model for phenotypic heterogeneity in breast cancer with implications for therapeutic strategies. *J.R.Soc. Interface* 2022, 18, 20210803.
16. Liu, R.; Wang, X.; Aihara, K.; Chen, L. Early diagnosis of complex diseases by molecular biomarkers, network biomarkers, and dynamical network biomarkers. *Med. Res. Rev*. **2014**, 34, 455-478.

17. Teschendorff, A. E.; Feinberg, A. P. Statistical mechanics meets single-cell biology. *Nat. Rev. Genet.* **2021**, 22, 459-476.
18. Chen, L.; Liu, R.; Liu, Z. P.; Li, M.; Aihara, K. Detecting early-warning signals for sudden deterioration of complex diseases by dynamical network biomarkers. *Sci. Rep*. **2012**, 2, 342-349.
19. Zhong, J.; Liu, R.; Chen, P. Identifying critical state of complex diseases by single-sample Kullback-Leibler divergence. BMC Genomics 2020, 21: 87
20. Scheffer, M. et al. Early-warning signals for critical transitions. *Nature* **2009**, 461, 53-59.
21. Thattai, M.; van Oudenaarden, A. Stochastic gene expression in fluctuating environments. *Genetics* **2004**, 167, 523-530.
22. Nevozhay, D.; Adams, R. M.; Van Itallie, E.; Bennett, M. R.; Balázsi, G. Mapping the environmental fitness landscape of a synthetic gene circuit. PLoS Comput. Biol. 2012, 8, e1002480.
23. Zhou, J. X.; Pisco, A. O.; Qian, H.; Huang, S. Nonequilibrium population dynamics of phenotype conversion of cancer cells. *PLoS ONE* **2014**, 9, e110714.
24. Goldenfeld, N. *Lectures on phase transitions and the renormalization group*, Frontiers in Physics 85; CRC Press: New York, USA, 2018; pp. 23-163.
25. Strogatz, S. H. *Nonlinear dynamics and chaos*, 1$^{st}$ ed.; Addision-Wesley: New York, USA, 1994; pp. 15-92.
26. Davies, P. C. W.; Demetrius, L.; Tuszynski T. A. Cancer as a dynamical phase transition. *Theoretical Biology and Medical Modelling*. **2011**, 8:30, 1-16.
27. Tsuchiya, M.; Giuliani, A.; Hashimoto, M.; Erenpreisa, J.; Yoshikawa, K. Emergent self-organised criticality in gene expression dynamics: temporal development of global phase transition revealed in a cancer cell line. *PLoS ONE* **2015**, 10, e0128565.
28. Tsuchiya, M.; Hashimoto, M.; Takenaka, Y.; Motoike, I. N.; Yoshikawa, K. Global genetic response in a cancer cell: self-organised coherent expression dynamics. *PLoS ONE* **2014**, 9, e97411.
29. Pal, M.; Ghosh, S.; Bose, I. Non-genetic heterogeneity, criticality and cell differentiation. Phys. Biol. 2015, 12, 016001.
30. Mojtahedi, M. et al. Cell fate decision as high-dimensional critical state transition. *PLoS Biol*. **2016**, 14, e2000640.
31. Cover, T. M.; Thomas, J. A. *Elements of information theory*, 2$^{nd}$ ed.; Wiley-Interscience: New Jersey, U. S. A., 2006; pp. 13-55.
32. Karolak, A.; Branciamore, S.; McCune, J. S.; Lee, P. P.; Rodin, S. S.; Rockne, R. C. Concepts and applications of information theory to immuno-oncology. *Trends in Cancer* **2021**, 7, 335-346.
33. Lin, J. Divergence measures based on the Shannon entropy. *IEEE Trans. Inf. Theory* **1991**, 37, 145-151.

34. Jinling, Y.; Peiluan, L.; Gao, R.; Ying, L; Luonan, C. Identifying critical states of complex diseases by single-sample Jensen-Shannon divergence. *Front. Oncol.* **2021**, 11, 684781.
35. Klein, I.; Doll, M. (Generalized) maximum cumulative direct, residual, and paired $\Phi$ entropy approach. Entropy 2020, 22, 91
36. Kauffman, S. Homeostasis and differentiation in random genetic control networks. *Nature* **1969**, 22, 177-178.
37. Huang, S.; Eichler, G.; Bar-Yam, Y.; Ingber, D. E. Cell fates as high-dimensional attractor states of a complex gene regulatory network. Phys. Rev. Lett. 2005, 94, 128701.
38. Li, Q. et al. Dynamics inside the cancer cell attractor reveal cell heterogeneity, limits of stability, and escape. *Proc. Natl. Acad. Sci. USA.* **2016**, 113, 2672-2677.
39. Kauffman, S. Differentiation of malignant to benign cells, *J. Theor. Biol*. **1971**, 3, 429-451.
40. Rockne, R. C. et al. State-transition analysis of time-sequential gene expression identifies critical points that predict development of acute myeloid leukemia. *Cancer Res.* **2020**, 80, 3157-3169.
41. Liu, R.; Yu, X.; Liu, X.; Xu, D.; Aihara, K.; Chen, L. Identifying critical transitions of complex diseases based on a single sample. Bioinformatics 2014, 30, 1579-1586.
42. Huang, S.; Kauffman, S.; How to escape the cancer attractor: Rationale and limitations of multi-target drugs. *Seminars in Cancer Biology* **2013**, 23, 270-278.
43. Dar, R. D.; Hosmane, N. N.; Arkin, M. R.; Siliciano, R. F.; Weinberger, L. S. Screening for noise in gene expression identifies drug synergies. *Science* **2014**, 344, 1392-1396.
44. Desai, R. V. et al. A DNA repair pathway can regulate transcriptional noise to promote cell-fate transitions. *Science* **2021**, 373, 870.